\documentclass[a4paper]{jpconf}
\usepackage{graphicx}
\usepackage{amsmath,amssymb,bm}
\usepackage{floatrow}
\newfloatcommand{capbtabbox}{table}[][\FBwidth]

\def \beq {\begin{equation}}
\def \edq {\end{equation}}
\def \bes {\begin{subequations}}
\def \eds {\end{subequations}}
\def \beqn {\begin{equation*}}
\def \edqn {\end{equation*}}

\def \dag {\dagger}

\def \up {\uparrow}
\def \down {\downarrow}

\def \calh {{\cal{H}}}

\begin{document}
\title{Nonlinear heat transport in ferromagnetic-quantum dot-superconducting systems}

\author{Sun-Yong Hwang$^{1}$ and David S\'{a}nchez$^2$}

\address{$^1$ Theoretische Physik, Universit\"at Duisburg-Essen and CENIDE, D-47048 Duisburg, Germany}
\address{$^2$ Institute for Cross-Disciplinary Physics and Complex Systems
IFISC (CSIC-UIB), E-07122 Palma de Mallorca, Spain}


\begin{abstract}
We analyze the heat current traversing a quantum dot sandwiched between a ferromagnetic
and a superconducting electrode. The heat flow generated in response to a voltage bias
presents rectification as a function of the gate potential applied to the quantum dot.
Remarkably, in the thermally driven case the heat shows a strong diode effect
with large asymmetry ratios that can be externally tuned with magnetic fields
or spin-polarized tunneling. Our results thus demonstrate the importance of hybrid
systems as promising candidates for thermal applications.
\end{abstract}

\section{Introduction}

Control of heat flow is a key goal in modern quantum electronics \cite{review,review3}.
Electrons carry energy in addition to charge and their transport can then be manipulated
electrically \cite{kul94,bog99,cip04,free06,zeb07,lei10,lop13b,whi13,hwa14,jia15,zim16}
or thermally \cite{wa07,seg08,ruo09,hwa13,san15,sie15}.
It is thus highly desirable to possess a great variety
of mesoscopic platforms where energy flow in response to various driving fields
can be generated and detected \cite{mol92,chi06,mes09,jez13}. 

Here, we investigate a ferromagnetic-quantum dot-superconducting (F-D-S) junction
and show that this system can work as an efficient heat diode both for charge and spin
transport. These hybrid systems have recently received a good deal of attention
due to their excellent thermoelectric properties \cite{oza14,kol16,kal14,mac14,hwa16a}.
Here, we show that they might also be attractive for thermal applications.


\section{Theory of nonlinear heat transport}
The total Hamiltonian describing the F-D-S system (left panel of Fig.~1) is given by~\cite{hwa16a}
\begin{equation}\label{eq:H}
\calh=\calh_{F}+\calh_{S}+\calh_{D}+\calh_{T}\,,
\end{equation}
where $\calh_{F}=\sum_{k\sigma}\varepsilon_{Fk\sigma}c_{Fk\sigma}^{\dag}c_{Fk\sigma}$
accounts for charge carriers in the ferromagnet with momentum $k$ and spin $\sigma=\up,\down$, and $\calh_{S}=\sum_{k\sigma}\varepsilon_{Sk\sigma}c_{Sk\sigma}^{\dag}c_{Sk\sigma}+\sum_{k}[\Delta c_{S,-k\up}^{\dag}c_{Sk\down}^{\dag}+{\rm H.c.}]$ depicts a superconductor reservoir with the energy gap $\Delta$ as an order parameter. Importantly, in the dot Hamiltonian
$\calh_{D}=\sum_{\sigma}(\varepsilon_{d\sigma}-eU_\sigma)d_{\sigma}^{\dag}d_{\sigma}$,
not only the energy level can be Zeeman split with magnetic fields, viz. $\varepsilon_{d\sigma}=\varepsilon_d+\sigma\Delta_Z$, but it can also be renormalized by the spin-dependent interaction potential $U_\sigma$. We determine $U_\sigma$ within a self-consistent (Hartree) approach~\cite{hwa14} to find the nonlinear heat transport in this setup. The tunneling between the dot and each lead is given by $\calh_{T}=\sum_{k\sigma}t_{F\sigma}c_{Fk\sigma}^{\dag}d_{\sigma}+\sum_{k\sigma}t_{S\sigma}c_{Sk\sigma}^{\dag}d_{\sigma}+{\rm H.c.}$,
which leads to broadenings $\Gamma_{F\sigma}=2\pi |t_{F\sigma}|^2 \sum_k \delta(\varepsilon-\varepsilon_{Fk\sigma})$
[parametrized as $\Gamma_{F\sigma}=\Gamma_F (1+\sigma p)$
with $p=(\nu_\up-\nu_\down)/(\nu_\up+\nu_\down)$ the $F$ polarization in terms of the density of states $\nu_\sigma=\sum_k \delta(\varepsilon-\varepsilon_{Fk\sigma})$ and $\Gamma_F=(\Gamma_{F\up}+\Gamma_{F\down})/2$]
and $\Gamma_{S\sigma}=\Gamma_S=2\pi |t_{S\sigma}|^2 \sum_p\delta(\varepsilon-\varepsilon_{Sp\sigma})$.

The spin-resolved heat current can be evaluated from the rate of energy flow per each spin ($\calh_{F\sigma}=\sum_{k}\varepsilon_{Fk\sigma}c_{Fk\sigma}^{\dag}c_{Fk\sigma}$) at the F side and the Joule heating in the presence of the voltage bias $V$
\beq
J_\sigma=-(i/\hbar)\langle[\calh,\calh_{F\sigma}]\rangle-I_\sigma V,
\edq
where the spin-resolved electric current is given by $I_\sigma=-(ie/\hbar)\langle[\calh,N_{F\sigma}]\rangle$ with the charge number $N_{F\sigma}=\sum_{k}c_{Fk\sigma}^{\dag}c_{Fk\sigma}$ for carriers with spin $\sigma$.
Due to coupling to the superconductor, the spin-resolved heat transport ($J_\sigma=J_{A}^{\sigma}+J_{Q}^{\sigma}$) has two separate contributions (with $J_{A}\equiv J_A^\up+J_A^\down$, $J_{A}^s\equiv J_A^\up-J_A^\down$, $J_{Q}\equiv J_Q^\up+J_Q^\down$, $J_{Q}^s\equiv J_Q^\up-J_Q^\down$)
\begin{align}
J_{A}^{\sigma}=\frac{-2eV}{h}&\int d\varepsilon~T_{A}^{\sigma}(\varepsilon)\big[f_{F}(\varepsilon-eV)-f_{F}(\varepsilon+eV)\big]=-2VI_{A}^{\sigma}\,,\label{eq:JA}\\
J_{Q}^{\sigma}=\frac{1}{h}&\int d\varepsilon~(\varepsilon-eV)~T_{Q}^{\sigma}(\varepsilon)\big[f_{F}(\varepsilon-eV)-f_{S}(\varepsilon)\big]\label{J_Q}\,,
\end{align}
where $J_{A}^{\sigma}$ refers to the spin-resolved Andreev heat current dominant for the subgap transport $|\varepsilon|<\Delta$, while $J_{Q}^{\sigma}$ is that of quasiparticle contributions beyond the gap $|\varepsilon|>\Delta$. Here, $f_{\alpha=F,S}(\varepsilon\pm eV)=\{1+\exp[(\varepsilon\pm eV-E_F)/k_{B}T_{\alpha}]\}^{-1}$ is the Fermi-Dirac distribution with the applied voltage to the ferromagnet $V=V_F$ and temperature $T_\alpha=T+\theta_\alpha$ ($T$: average temperature, $\theta_F=\theta$: thermal bias, $V_S=\theta_S=0$). The respective transmission functions $T_{A}^{\sigma}(\varepsilon)$ and $T_{Q}^{\sigma}(\varepsilon)$ are evaluated using the Green's function approach
(see, e.g., Refs.~\cite{hwa16a,cao04,hwa17} for the explicit expressions).

In a previous work~\cite{hwa16a}, we reported a large thermopower in the same setup but only linear response was considered.
Thus, the potential shift $U_\sigma$ in $\calh_D$ of Eq.~\eqref{eq:H} could be neglected. Further, the linear Andreev heat current was shown to
be zero. Here, we consider the nonlinear case, where $U_\sigma$ is determined in a weakly nonequilibrium condition~\cite{hwa17}.
In Eq.~\eqref{eq:JA} the Andreev electric current reads $I_A^\sigma=V(G_{A0}^\sigma+G_{A1}^\sigma V+M_{A}^\sigma \theta+\dots)$ (purely thermoelectric terms  vanish due to the particle-hole symmetry~\cite{hwa15}). Then, the Andreev heat flow becomes
\beq\label{eq:JA2}
J_{A}^{\sigma}=-2V^2(G_{A0}^\sigma+G_{A1}^\sigma V+M_{A}^\sigma \theta+\dots)\,.
\edq
Note that the leading order nonvanishing Andreev thermal conductance can only be given by the cross coupling term $M_A^\sigma$ similarly to the subgap nonlinear electric current~\cite{hwa15}. This implies that in the isoelectric case $V=0$ the subgap thermal transport is entirely blocked and the thermal heat current will be activated by the quasiparticle contributions only. This effect has exactly the same origin as the recently proposed Seebeck diodes~\cite{hwa16b}. However, if we apply a finite voltage bias $V$ the subgap heat current compete with the quasiparticle contributions and hence  Andreev-Peltier effects can become important. This will be discussed below for pure isothermal cases. Finally, if we apply high enough thermal gradient for a nonzero $V$, quasiparticles dominantly contribute to the heat transport after the competing regime is over where $J_Q\simeq-J_A$ and $J_Q^s\simeq-J_A^s$ (see Fig.~3). Beyond this competing regime, large heat and spin heat currents can be generated in our device from quasiparticle tunneling.

\begin{figure}[t]
\centering
\includegraphics[width=0.3\textwidth, height=6cm,clip]{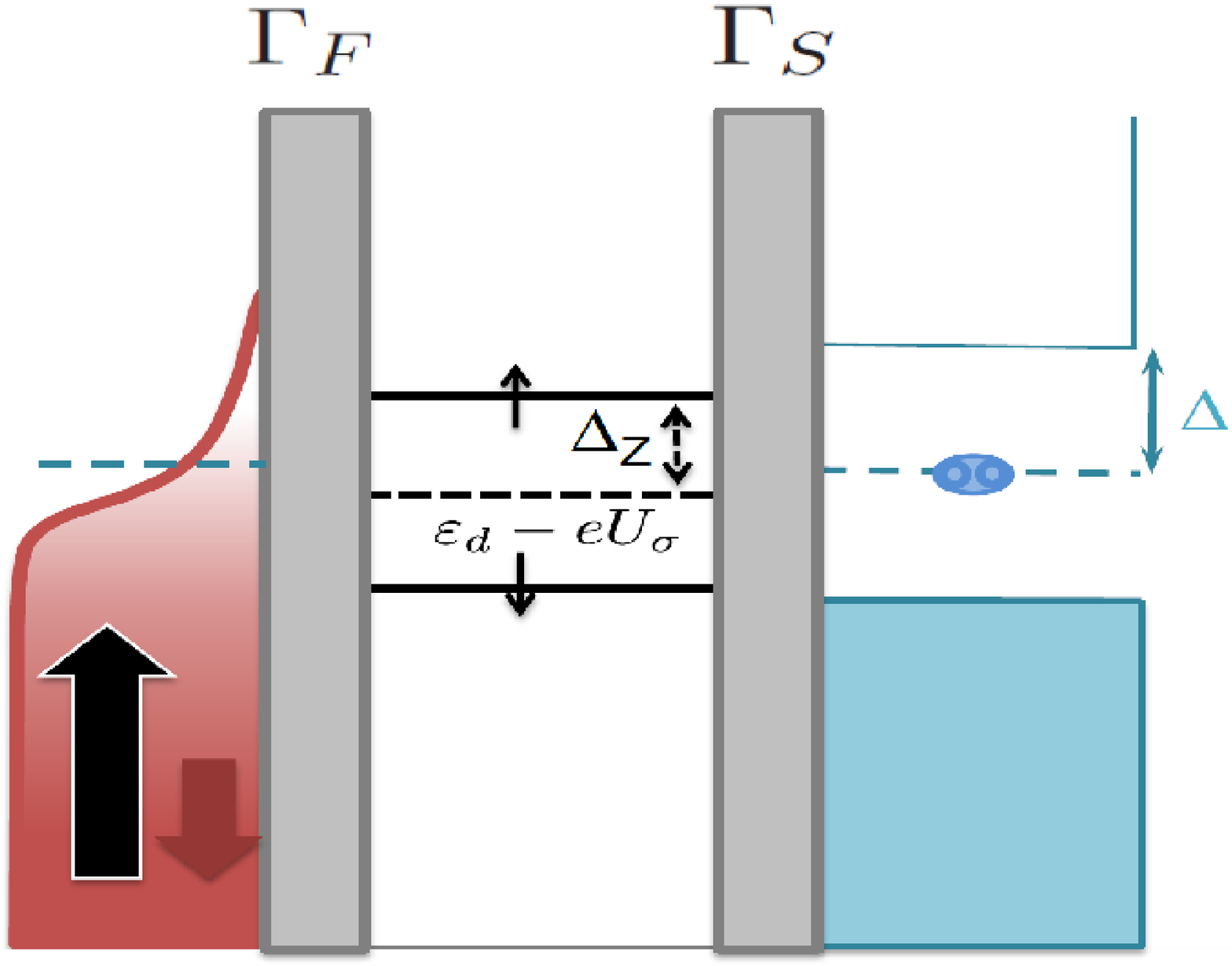}
\includegraphics[width=0.65\textwidth, clip]{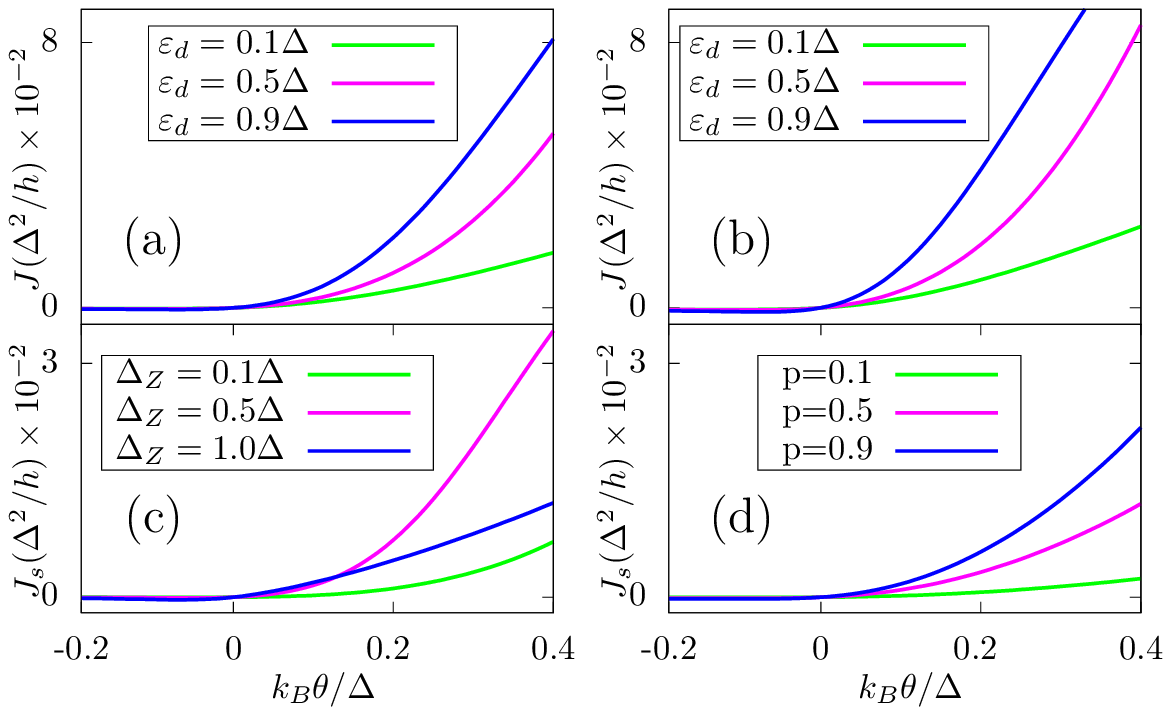}
\caption{Left panel: sketch of our F-D-S device energy diagram. The superconducting reservoir shows an energy gap $\Delta$. The left ferromagnet has a spin polarization $p$ with different amounts of up and down spin component generating spin-dependent tunneling rates. $\Gamma_{F,S}$ is the spin-averaged tunnel broadening to the quantum dot from F or S contact. Thermal and electric biases are applied only to the ferromagnet. The dot level $\varepsilon_d$ depends on $\sigma=\up,\down$ due to either Zeeman splitting $\Delta_Z$ or Coulomb potential $U_\sigma$, which is a self-consistent function of voltage and temperature biases. Right panel:
thermal [(a),(b)] and spin thermal [(c),(d)] diode effects of our device ($V=0$). Charge heat flux for (a) $\Gamma_F=0.1\Delta$, $\Gamma_S=0.5\Delta$, (b) $\Gamma_F=0.5\Delta$, $\Gamma_S=0.1\Delta$, and spin heat current at $\varepsilon_d=0.3\Delta$ with $\Gamma_F=0.1\Delta$, $\Gamma_S=0.5\Delta$ for (c) $p=0$, (d) $\Delta_Z=0$ are shown as a function of thermal bias $\theta$. Base temperature is $k_BT=0.2\Delta$.
}\label{fig2}
\end{figure}

\begin{table}
\begin{center}
\begin{tabular}{|l|c|c|c|}
\hline
& $k_B\theta_0=0.05\Delta$ & $k_B\theta_0=0.10\Delta$ & $k_B\theta_0=0.15\Delta$ \\
\hline
(a) $\varepsilon_d=0.5\Delta$ & $3.25$ & $9.41$ & $21.6$\\
\hline
(b) $\varepsilon_d=0.5\Delta$ & $3.02$ & $8.34$ & $18.7$\\
\hline
(c) $\Delta_Z=0.5\Delta$ & $4.71$ & $19.9$ & $67.5$\\
\hline
(d) $p=0.5$ & $2.89$ & $7.46$ & $15.3$\\
\hline
\end{tabular}\label{tab}
\end{center}\caption{Asymmetry ratio $R=\frac{|J_{(s)}(\theta_0)|}{|J_{(s)}(-\theta_0)|}$ with denoted parameter values from Fig.~1.}
\end{table}

Having the aforementioned points in mind, we below discuss the total heat flux $J=J_\up+J_\down$ and the spin-polarized heat current $J_s=J_\up-J_\down$ where $J_\sigma=J_{A}^{\sigma}+J_{Q}^{\sigma}$.

\section{Results and discussion}
\subsection{Thermal diode effects}
Figure 1 shows the thermal diode effects appearing in the isoelectric case ($V=0$) where subgap heat transport is completely blocked. This device thus provides means to control the heat flow in a unidirectional way. This is akin to the diode effects of the thermoelectric currents~\cite{hwa16b}. Moreover, the spin polarized quasiparticle heat currents can be rectified if spin symmetry is broken by magnetic fields or coupling to the F lead [Figs.~1(c) and 1(d)].
In Figs.~1(a) and 1(b), the heat flux signal is slightly larger in F-dominant case [(b) $\Gamma_F > \Gamma_S$] but the rectification efficiency is higher in S-dominant case [(a) $\Gamma_F < \Gamma_S$], as shown in Table 1 with the asymmetry ratio $R$ at $\varepsilon_d=0.5\Delta$. In Fig.~1, we use the average temperature $k_BT=0.2\Delta$ but the efficiency quickly increases as we lower $T$. At $k_BT=0.1\Delta$, for example, $R$ in Table 1 increases for (a) 3.25 $\to$ 49.7 and (b) 3.02 $\to$ 43 at $k_B\theta_0=0.05\Delta$.
In Fig.~1(c), the spin heat current for $\theta>0$ (say at $k_B\theta=0.2\Delta$) displays a nonmonotonic dependence on the Zeeman splittings as it firstly increases with $\Delta_Z$ and can be maximized and then decreases again, because the dot level splitting by $\Delta_Z$ generates a strong energy dependence of the heat transport. Consequently, a strong spin heat rectification is possible in this optimal Zeeman splitting as shown in Table 1 at $k_B\theta_0=0.15\Delta$ where $R\simeq70$. This, however, is not the case for the spin asymmetry created solely from the F polarization $p$, which monotonously enhances the spin heat flow as $p$ increases [Fig.~1(d)].


\subsection{Nonlinear Peltier effect}
\begin{figure}[t]
\centering
\includegraphics[width=0.65\textwidth, clip]{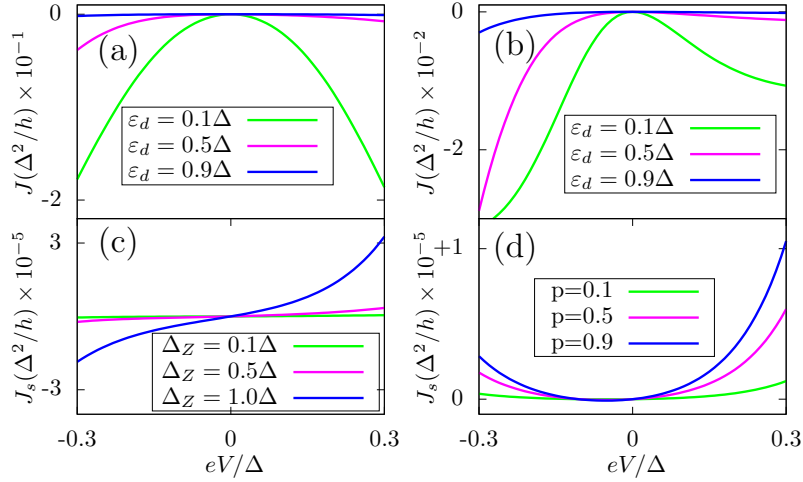}
\caption{Nonlinear Peltier effects of our device in the isothermal case $\theta=0$. Charge heat flux for (a) $\Gamma_F=0.1\Delta$, $\Gamma_S=0.5\Delta$, (b) $\Gamma_F=0.5\Delta$, $\Gamma_S=0.1\Delta$, and spin heat current at $\varepsilon_d=0.3\Delta$ with $\Gamma_F=0.1\Delta$, $\Gamma_S=0.5\Delta$ for (c) $p=0$, (d) $\Delta_Z=0$ are shown as a function of the voltage $V$. Background temperature is $k_BT=0.1\Delta$.
}\label{fig2}
\end{figure}

An applied voltage bias $V$ is detrimental to the diode effects described above as the subgap heat transport becomes appreciable beyond the linear response [Eq.~\eqref{eq:JA2}]. Nonetheless, one can achieve nonlinear
effects~\cite{lop13b} thanks to this voltage bias. In Fig.~2, the charge heat [(a) $\Gamma_F<\Gamma_S$, (b) $\Gamma_F>\Gamma_S$] and the spin heat [either (c) $\Delta_Z\ne0$ or (d) $p\ne0$] fluxes are displayed in the isothermal case ($\theta=0$). The S-dominant case in Fig.~2(a) has much larger heating currents (about one order of magnitude) than the opposite coupling limit shown in Fig.~2(b). At optimal gate potential, e.g., $\varepsilon_d=0.5\Delta$, the heat can be rectified. Our device can thus act as Peltier diodes and the asymmetry ratio $R=\frac{|J(-V_0)|}{|J(V_0)|}$ for a given $V_0>0$ is much larger in the F-dominant case as displayed in (b). However, this diode effect tends to be fragile 
because the Andreev Joule heating is now manifested over the subgap energy range ($|\varepsilon|<\Delta$).
As shown in Fig.~2(c), the sign of the spin polarized heat flux can be positive or negative depending on the bias voltage direction when the spin asymmetry is generated by $\Delta_Z$ in contrast to the asymmetry due to $p$ [Fig.~2(d)]. Notice that the voltage-driven spin heat flows in Figs.~2(c),(d) are several orders of magnitude smaller than the voltage-driven total heat flux [Figs. 2(a),(b)] or the (positive) temperature-driven spin heat flows [Figs. 1(c),(d); for direct comparisons we also generate same figures at $k_BT=0.1\Delta$ (not shown) that give smaller values than those at $k_BT=0.2\Delta$ but within the same order of magnitude]. Therefore, one needs to apply high enough thermal gradients even with finite voltages in order to activate spin-polarized quasiparticles. A quick understanding of this can be supplied by observing Figs. 3(b) and 3(c), where one can find $J_Q^s\simeq-J_A^s$
below $\theta\simeq T$ hence $J_s=J_A^s+J_Q^s\simeq0$ albeit the applied voltage $eV=k_BT$. Now we will further discuss the results in Fig.~3.


\begin{figure}[t]
\centering
\includegraphics[width=1.0\textwidth, clip]{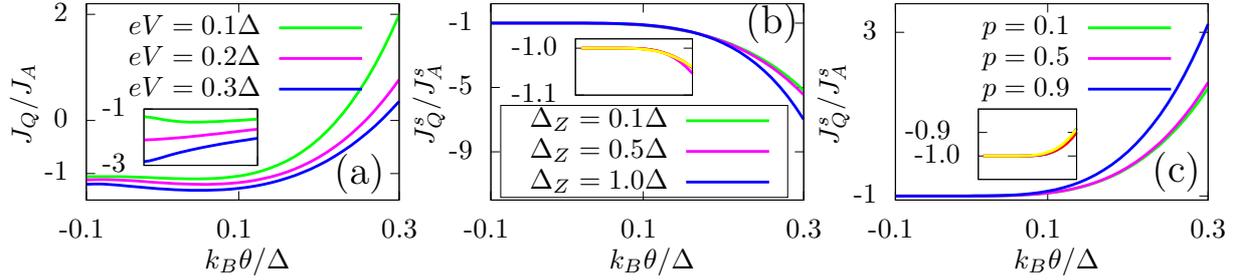}
\caption{Competition with the subgap Joule heating in generic thermoelectric configurations. We use $k_BT=\Gamma_F=0.1\Delta$, $\Gamma_S=\varepsilon_d=0.5\Delta$, and $eV=0.1\Delta$ in (b) $p=0$ and (c) $\Delta_Z=0$. The competing regime can be identified with $J_Q\simeq-J_A$ and $J_Q^s\simeq-J_A^s$ canceling the total heat currents. Each inset is explained in the main text.
}\label{fig2}
\end{figure}

\subsection{Competition between Andreev and quasiparticle contributions}
It should be again noted that there is no linear subgap Peltier effect in our system due to Onsager symmetry since there is no corresponding linear Andreev thermocurrent because of the particle-hole symmetry~\cite{hwa15}. Indeed, in Eq.~\eqref{eq:JA} the energy carried by particles and holes exactly cancel each other and only the Joule heating remains~\cite{hwa16a}. The latter $J_A(J_A^{s})$ which appears only beyond the linear response can now compete with the normal heat flow $J_Q(J_Q^{s})$ carried by quasiparticles. Hence, $J_A(J_A^{s})$ can create the tendencies of cooling with positive voltages while at the same time $J_Q(J_Q^{s})$ tends to heat, or vice versa.

In Fig.~3(a), one finds $J_Q\simeq-J_A$ for a low voltage bias, e.g., $eV=0.1\Delta$ below $k_B\theta\simeq0.1\Delta= k_BT$ (the actual value is about $J_Q=-1.06J_A$). In this region, we have a very small total flux since $J=J_A+J_Q\simeq-0.06J_A\simeq0$. But as $V$ is increased, quasiparticles slightly dominate and at $eV=0.3\Delta$ we have $J_Q\simeq-1.2J_A$. At $\theta=0$, this can be compared with Fig.~2(a) at $\varepsilon_d=0.5\Delta$, in which one can notice a tendency of $J$ to increase (as $V>0$ is applied) to negative values since $J\simeq-0.2J_A<0$. A small detuning of the dot level is more beneficial for this purpose as shown in Fig. 2(a) with $\varepsilon_d=0.1\Delta$. Inset of Fig. 3(a) shows the corresponding curves in the range $-0.1\Delta<k_B\theta<0.1\Delta$ but with $\varepsilon_d=0.1\Delta$. Indeed, one can notice an appreciable contribution from quasiparticles as the voltage is applied reaching $J_Q\simeq-2.4J_A$ at $eV=0.3\Delta$ and $\theta=0$, which explains the behavior in Fig. 2(a).
As thermal gradient further increases, see Fig.~3(a), a compensating regime appears where $J_Q=0$ due to the combined thermoelectric configurations (a voltage-driven quasiparticle heat compensates for the temperature-driven flow), after which $J_Q$ tends to collaborate with $J_A$ giving rise to a net heating. At $eV=0.1\Delta$ and $k_B\theta=0.3\Delta$, for instance, $J_Q\simeq 2J_A$ and hence the total flux is given by $J\simeq3J_A$.

In Fig.~3(b), when the spin heat flow is generated by magnetic fields, there is no compensating or collaborating regime but the amplitude of $J_Q^s$ keeps increasing against $J_A^s$ reaching $J_Q^s\simeq-5J_A^s$ at $k_B\theta=0.3\Delta$ and $\Delta_Z=eV=0.1\Delta$. However, the polarization-driven spin heat currents in Fig. 3(c) exhibit the similar tendency with separate regimes as the curves in Fig. 3(a), i.e., crossing the zero and eventually becoming positive.
Finally, the insets of Figs.~3(b) and 3(c) show the voltage dependence of the curves in the range $-0.1\Delta<k_B\theta<0.1\Delta$. The red and yellow lines respectively refer to $eV=0.2\Delta$ and $eV=0.3\Delta$ at (b) $\Delta_Z=0.5\Delta$, and (c) $p=0.9$. In stark contrast to the voltage dependence of the total heat flux in Fig. 3(a), the competing effects are robust with respect to the applied $V$ in these spin heat cases, maintaining $J_Q^s\simeq-J_A^s$. This explains the vanishingly small amplitudes of $J_s$ in Figs. 2(c) and 2(d) as one specific example of the isothermal case $\theta=0$. Indeed, in a broad range of $-0.1\Delta<k_B\theta<0.06\Delta$, the voltage dependence is very small calling for a substantial amount of thermal gradients to observe the spin polarized heat in a general thermoelectric bias configuration.

It should be emphasized that the above discussions are meaningful only with the finite voltage $V\ne0$, otherwise $J_A=J_A^{s}=0$ [Eq.~\eqref{eq:JA2}] and the ratios $J_Q/J_A$ and $J_Q^{s}/J_A^{s}$ in Fig.~3 will diverge. Since the discussed effects are intrinsically nonlinear, the competition between $J_A$ and $J_Q$ ($J_A^{s}$ and $J_Q^{s}$) appears even for small driving fields. This is completely different from the case for the electric currents, where one should have $I_A\gg I_Q\simeq0$ for a low bias regime and the competition between the Andreev and quasiparticle parts in this case will be meaningless.

\section{Conclusions}
To summarize, we have examined the heat transport properties of a quantum dot attached to ferromagnetic
and superconducting contacts. We have unveiled strong thermal diode effects that can be manipulated
with external magnetic fields or magnetization of the ferromagnetic reservoir.
In these systems, it is crucial to distinguish between Andreev and quasiparticle currents, which determine
the specific transport mechanism. Our work thus represents an important step toward a full characterization
of heat transport in hybrid systems.

\ack
This work was supported by MINECO under Grant  No.\ FIS2014-52564 and the Ministry of Innovation NRW.

\end{document}